\documentclass[12pt,a4paper]{article}
\usepackage{epsf}
\begin{document}
\title{Detecting top-Higgs at high energy
$e^{+}e^{-}$ colliders}
\author{Chongxing Yue$^{(a,b)}$,Guoli Liu$^{b}$,
 Qingjun Xu$^{b}$\\ {\small a: CCAST (World
Laboratory) P.O. BOX 8730. B.J. 100080 P.R. China}
\\ {\small b:College of Physics and Information Engineering,}\\
\small{Henan Normal University, Xinxiang  453002. P.R.China}
\thanks{This work is supported by the National Natural Science
Foundation of China(I9905004), the Excellent Youth Foundation of
Henan Scientific Committee(9911); and Foundation of Henan
Educational Committee.}
\thanks{E-mail:cxyue@public.xxptt.ha.cn} }
\date{\today}
\maketitle
\begin{abstract}
\hspace{5mm}We calculate the contributions of the top-Higgs
$h_{t}^{0}$ predicted by topcolor assisted technicolor(TC2) models
to $e^{+}e^{-}\longrightarrow
\overline{t}c\overline{\nu}_{e}\nu_{e}$ and compare the results
with the contributions of $h^{0}_{t}$ to the processes
$e^{+}e^{-}\longrightarrow Z h^{0}_{t}\longrightarrow
Z\overline{t}c$ and $e^+e^-\longrightarrow \gamma h^{0}_{t}
\longrightarrow \gamma \overline{t}c$. We find that $e^{+}e^{-}
\longrightarrow\overline{t}c\overline{\nu}_{e}\nu_{e}$ is very
sensitive to $h^{0}_{t}$, which can be easy detected via this
process at high-energy $e^{+}e^{-}$ collider(LC) experiments with
$\sqrt{s}\geq 500$ $GeV$, as long as its mass below the
$\overline{t}t$ threshold. The process $e^{+}e^{-} \longrightarrow
\gamma\overline{t}c$ also can be used to detect $h^{0}_{t}$ at LC
experiments.
\end {abstract}

\vspace{1.0cm} \noindent
 {\bf PACS number(s)}: 12.60Fr, 14.80.Cp, 12.60.Rc

\newpage
     An important issue in high-energy physics is to understand
the mechanism of the mass generation. The top quark, with a mass
of the order of the weak scale, is singled out to play a key role
in the dynamics of the electroweak symmetry breaking (EWSB) and
flavor symmetry breaking. There may be a common origin for EWSB
and top quark mass generation. Much theoretical work has been
carried out in connection to the top quark and EWSB. The
topcolor-assisted technicolor (TC2) models\cite{y1}, the top
see-saw models\cite{y2} and the flavor universal coloron
models\cite{y3} are three of such examples. Such type of models
generally predict a number of scalars with large Yukawa couplings
to the third generation. For example, TC2 models predict the
existence of the top-pions ($\pi^{\pm}_{t}$,$\pi^{0}_{t}$) and the
neutral CP-even state, called top-Higgs $h^{0}_{t}$, which is
analogous to the $\sigma$ particle in low energy QCD. These new
particle are most directly related to the EWSB. Thus, studying the
possible signatures of these new particles at high energy
colliders would provide crucial information for the EWSB and
hopefully fermion flavor physics as well.

    It is well known that there is no flavor changing neutral
current (FCNC) at tree-level in the standard model (SM). The
production cross section of the FCNC process is very small at
one-loop level due to the unitary of CKM matrix. Thus, the FCNC
process can be used to search for new physics. Any observation of
the flavor changing coupling deviated from that in the SM would
unambiguously signal the presence of new physics. For TC2
models\cite{y1}, the underlying interactions, topcolor
interactions, are non-universal and therefore do not possess a GIM
mechanism. The scalars predicted by this kind of models can induce
the flavor changing scalar couplings and give distinct new flavor
mixing phenomena which may be tested at both low and high energy
experiments\cite{y4,y5}. In this paper, we calculate the
contributions of the top-Higgs $h^{0}_{t}$ to the t-channel vector
boson fusion process $e^{+}e^{-} \longrightarrow
W^{+}W^{-}\overline{\nu}_{e}\nu_{e}\longrightarrow
\overline{t}c\overline{\nu}_{e}\nu_{e}$ and compare with the
contributions of $h^{0}_{t}$ to the processes $e^{+}e^{-}
\longrightarrow Zh^{0}_{t} \longrightarrow Z\overline{t}c$ and
$e^+e^-\longrightarrow \gamma h^{0}_{t} \longrightarrow \gamma
\overline{t}c$. Our results show that the top-Higgs $h^{0}_{t}$
can give significantly contributions to the process
$e^{+}e^{-}\longrightarrow \overline{t}c\overline{\nu}_{e}\nu_{e}$
which can be detected at the high-energy $e^{+}e^{-}$ colliders
(LC) with the center-of-mass energy $\sqrt{s}=500-1500$ $GeV$, as
long as its mass is below $2m_{t}$. The effects of $ h_{t}^{0} $
on the process $e^+e^-\longrightarrow  \gamma \overline{t}c$ also
can be detected at the LC experiments.

      According the idea of the TC2 models, there is the following
relation:
\begin{equation}
\nu^{2}_{\pi}+F^{2}_{t}=\nu^{2}_{w},
\end{equation}
where $\nu_{\pi}$  represents the contributions of the TC or other
interactions to the EWSB, $F_{t}\simeq 50$ $GeV$ is decay constant
of the scalars (top-pions or top-Higgs) predicted by the TC2
models, and $\nu_{w}=v/\sqrt{2}\simeq 174$ $GeV$. Thus, the
majority of masses of gauge bosons $W$ and $Z$ come from the
technifermion condensate. The couplings of the top-Higgs
$h^{0}_{t}$ to the electroweak  gauge bosons $W$ and $Z$ at tree
level are suppressed by the factor $F_{t}/\nu_{w}$ with respect to
that of the SM Higgs H:

\begin{equation}
h^{0}_{t}WW:\frac{F_{t}}{\nu_{w}}gm_{W}g_{\mu\nu},\hspace{5mm}
h^{0}_{t}ZZ:\frac{F_{t}}{\nu_{w}}\frac{gm_{Z}}{\cos\theta_{w}}g_{\mu\nu}.
\end{equation}
The couplings of the top-Higgs $h^{0}_{t}$ to the third generation
quarks, including the $t-c$ transition, can be written
as\cite{y1,y4}:
\begin{equation}
h^{0}_{t}\overline{t}t:\frac{m_{t}}{\sqrt{2}F_{t}}
\frac{\sqrt{\nu_{w}^{2}-F_{t}^{2}}}{\nu_{w}}K^{tt},\hspace{5mm}
h^{0}_{t}\overline{t}c:\frac{m_{t}}{\sqrt{2}F_{t}}
\frac{\sqrt{\nu_{w}^{2}-F_{t}^{2}}}{\nu_{w}}K^{tc},\hspace{5mm}
h^{0}_{t}\overline{b}b:\frac{m_{b}^{*}}{\sqrt{2}F_{t}}
\frac{\sqrt{\nu_{w}^{2}-F_{t}^{2}}}{\nu_{w}}.
\end{equation}
$m_{b}^{*}$ is the part of the bottom quark mass generated by
instanton effects, which we assume $m_{b}^{*}=0.8$$m_{b}$. It has
been shown\cite{y4} that $ K^{tt}=1-\epsilon $ and
$K^{tc}=\sqrt{(K_{UR}^{tc})^2+(K_{UL}^{tc})^2}\simeq K_{UR}^{tc}
\leq\sqrt{\epsilon-\epsilon^{2}}$, which $\epsilon$ is a model
dependent parameter. In this paper, we assume that the part of the
top quark mass generated by the topcolor interactions makes up
$99\%$ of $m_{t}$, i.e. $\epsilon=0.01$ and take the parameter
$K^{tc}$ as a free parameter.

   The couplings of the top-Higgs $h^{0}_{t}$ to gauge boson pairs
gg, $\gamma\gamma$ or $Z\gamma$ are similar to those of the
neutral top-pion $\pi^{0}_{t}$ which come from the top quark
triangle loop. The general form has been given in Ref.[5,6]. Thus,
for $200GeV\leq m_{h_{t}}\leq 400GeV$, the total decay width
$\Gamma$ mainly comes from the decay modes $b\overline{b}$,
$\overline{t}c$, $WW$, $ZZ$, $gg$ and $t\overline{t}$ (if
kinematically allowed).

     The process $e^{+}e^{-}\longrightarrow\overline{t}c
\overline{\nu}_{e}\nu_{e}$ can be well approximated by the $
W^{+}W^{-}$ fusion process:
$W^{+}W^{-}\longrightarrow\overline{t}c$. It has been
shown\cite{y7} that the effective W-boson approximation (EWA)
provides a viable simplification for high energy processes
involving $W^{+}W^{-}$ fusion. Thus, we use the effective EWA to
calculate the contributions of the top-Higgs $h^{0}_{t}$ to the
process $e^{+}e^{-}\longrightarrow
\overline{t}c\overline{\nu_{e}}\nu_{e}$ and discuss the
observability of $h_{t}^{0}$ at the LC experiments with $
\sqrt{s}=500GeV-1500GeV$.

    The production cross section of the subprocess
$W^{+}_{\lambda_{+}}W_{\lambda_{-}}^{-}\longrightarrow\overline{t}c$
generated by the top-Higgs $h^{0}_{t}$ can be written as:
\begin{equation}
\widehat{\sigma}(W^{+}_{\lambda_{+}}W_{\lambda_{-}}^{-}\longrightarrow
\overline{t}c)=\frac{N_{c}\alpha}{4S_{w}^{2}}
\frac{\nu_{w}^{2}-F_{t}^{2}}{\nu_{w}^{4}}
(K^{tc})^{2}|\epsilon_{\lambda_{+}}^{W^{+}}\cdot
\epsilon_{\lambda_{-}}^{W^{-}}|^{2} \frac{m_{W}^{2}m^{2}_{t}}
{(\widehat{s}-m_{h_{t}}^{2})^{2}+m_{h_{t}}^{2}\Gamma^{2}} \cdot
\frac{\beta_{t}^{4}}{\beta_{W}},
\end{equation}
with
\begin{equation}
\beta_{t}=\sqrt{1-\frac{m_{t}^{2}}{\widehat{s}}},\hspace{5mm}
\beta_{W}=\sqrt{1-\frac{4m_{W}^{2}}{\widehat{s}}}.
\end{equation}
Where $\Gamma$ is the total width of $h_{t}^{0}$ and $
\sqrt{\widehat{s}}$ is the  center-mass-energy of the WW
center-mass frame. Due to a severe CKM suppression, the cross
section $ \widehat{\sigma}(W_{\lambda_{+}}^{+}W^{-}_{\lambda_{-}}
\longrightarrow \overline{t}c)$ is very small in the SM:
$\sigma_{SM}^{\overline{t}c\overline{\nu}\nu}\sim
10^{-5}-10^{-4}$fb for $ \sqrt{s}=500-2000GeV$\cite{y8}. Thus, in
above formula, we have neglected the SM contributions.

  The cross section $\sigma^{\overline{t}c\overline{\nu}\nu}$ of
the process $e^{+}e^{-}\longrightarrow
\overline{t}c\overline{\nu}_{e}\nu_{e}$ can be obtained by folding
the cross section $
\widehat{\sigma}(W_{\lambda_{+}}^{+}W^{-}_{\lambda_{-}}
\longrightarrow \overline{t}c)$ with the distribution functions
$f^{W}_{\lambda_{i}}$.
\begin{equation}
\sigma^{\overline{t}c\overline{\nu}\nu}=\sum_{\lambda_{+},\lambda_{-}}
\int\int dx_{+}dx_{-}
f_{\lambda_{+}}(x_{+})f_{\lambda_{-}}(x_{-})\widehat{\sigma}
(W_{\lambda_{+}}^{+}W_{\lambda_{-}}^{-} \longrightarrow
\overline{t}c),
\end{equation}
where the helicities $\lambda_{\pm}$ of the $W^{\pm}$ each run
over 1, 0, -1. $f_{\lambda_{+}}(x_{+})$ and
$f_{\lambda_{-}}(x_{-})$ are the distribution functions of $W^{+}$
and $W^{-}$, respectively. Similarly to Ref.[8], we use the full
distribution functions given by Ref.[7] and include all
polarizations for the W boson in our calculations.

   In Fig.1, we plot the cross section $\sigma^{\overline{t}c
\overline{\nu}\nu}$ as a function of $m_{h_{t}}$ for $
\sqrt{s}=500GeV$ and four  values of the parameter $K^{tc}$. From
Fig.1 we can see that $\sigma^{\overline{t}c\overline{\nu}\nu}$
increases with increasing the value of $ K^{tc} $ and reach the
maximum value for $m_{h_{t}}=215GeV$. For $m_{h_{t}}>2m_{t}$, the
$\sigma^{\overline{t}c \overline{\nu}\nu}$ drops considerably
since the $ \overline{t}t$ channel opens up and the branching
ratio $B_{r}(h_{t}^{0}\longrightarrow \overline{t}c)$ drops
substantially. If we assume that a yearly integrated luminosity of
a LC experiment with $ \sqrt{s}=500GeV$ is about $50fb^{-1}$, then
the cross section $\sigma^{\overline{t}c \overline{\nu}\nu}$ would
yield several tens to hundreds such events in most of the
parameter space ($m_{h_{t}}$,$K^{tc}$). For example, for
$m_{h_{t}}=300GeV$ and $K^{tc}=0.05$, there would be $25$ such
events to be generated at the LC experiments with $
\sqrt{s}=500GeV$.

   It is well known that, for high energy process involving
$W^{+}W^{-}$ fusion, the corresponding cross section grows with
the center-mass-energy $ \sqrt{s}$ of colliders. To see the
effects of $\sqrt{s}$ on the production cross section, we plot the
cross section  $\sigma^{\overline{t}c \overline{\nu}\nu}$ versus $
m_{h_{t}} $ in Fig.2 for $K^{tc}=0.05$ and four values of
$\sqrt{s}$. The cross section increases from $1.24fb$ to $9.97fb$
as $ \sqrt{s}$ increases from $500GeV$ to $ 1500GeV$ for
$m_{h_{t}}=250GeV$. It is evident from Fig.2 that the top-Higgs
$h_{t}^{0}$ would produce several hundreds to thousands of
$\overline{t}c\overline{\nu}_{e}\nu_{e}$ events in a LC running at
$ \sqrt{s}\geq 1000GeV$ with an integrated luminosity of $L\geq
100$$fb^{-1}$. Thus, it is very easy to detect the signals of
$h_{t}^{0}$ via this process at a LC experiment with $
\sqrt{s}\geq 1000GeV$.

  Certainly, the top-Higgs $h_{t}^{0}$ has contributions to the
process $e^{+}e^{-}\longrightarrow \overline{t}ce^{+}e^{-}$ via
the subprocess $ZZ\longrightarrow h_{t}^{0}\longrightarrow
\overline{t}c$. The main difference between $\sigma^{\overline{t}c
\overline {\nu} \nu} $ and $\sigma^{\overline{t}cee}$ arises from
the dissimilarity between the distribution functions for W and Z
bosons. Such the distribution functions of gauge boson Z are
smaller than those of gauge boson W, $\sigma^{\overline{t}cee}$ is
expected to be smaller by about one order of magnitude than
$\sigma^{\overline{t}c \overline{\nu}\nu}$\cite{y8}. This
conclusion is model independent. Thus, we do not consider the
contributions of the top-Higgs $h_{t}^{0}$ to the process
$e^{+}e^{-}\longrightarrow \overline{t}ce^{+}e^{-}$ in this paper.

   The top-Higgs  $h_{t}^{0}$ also has contributions to the
process $e^{+}e^{-}\longrightarrow Z\overline{t}c$ via the Bjorken
process $e^{+}e^{-}\longrightarrow Zh_{t}^{0}$ followed by
$h_{t}^{0}\longrightarrow\overline{t}c$. Thus, $h_{t}^{0}$ may be
detected via this process at the LC experiments. To compare the
production rate of
$e^{+}e^{-}\longrightarrow\overline{t}c\overline{\nu}_{e}\nu_{e}$
with that of $e^{+}e^{-}\longrightarrow Z\overline{t}c$ in TC2
models, we plot the rate
$R_{1}=\frac{\sigma^{\overline{t}c\overline{\nu}\nu}}
{\sigma^{Z\overline{t}c}}$ as a function of $m_{h_{t}}$ for
$K=0.05$ and three values of $ \sqrt{s}$ in Fig.3. We find that
the cross section $\sigma^{Z\overline{t}c}$ decreases with
increasing the values of $m_{h_{t}}$ or $\sqrt{s}$. For
$\sqrt{s}\geq 500GeV$, the ratio $R_{1}$ is larger than 1 in most
of the parameter space. For example, for $m_{h_{t}}=300GeV $, $
K^{tc}=0.05$ and $\sqrt{s}= 500GeV $ , the ratio $R_{1}$
approximately equals to $ 2$. Thus, it is very difficult to detect
the top-Higgs $h_{t}^{0}$ via the process
$e^{+}e^{-}\longrightarrow Z\overline{t}c $ at LC experiments with
$\sqrt{s}\geq 500GeV$.

   Since the top-Higgs $h_{t}^{0}$ can couple to gauge boson pair
$ VV $ via the quark triangle loop, the top-Higgs may have
significantly contributions to process $ e^{+}e^{-}\longrightarrow
h_{t}^{0}\gamma \longrightarrow \gamma\overline{t}c $. This has
been studied in Ref.[5] for the neutral top-pion $\pi_{t}^{0}$. To
see whether $h_{t}^{0}$ can be detected via
$e^{+}e^{-}\longrightarrow\gamma\overline{t}c $, we plot the ratio
$R_{2}=\frac{\sigma^{\overline{t}c\overline{\nu}\nu}}
{\sigma^{\gamma\overline{t}c}}$ as a function of $m_{h_{t}}$ for
$K^{tc}=0.05$ and three values of $ \sqrt{s}$ in Fig.4. From
Fig.4, we can see that the cross section
$\sigma^{\gamma\overline{t}c}$ increases with $ \sqrt{s}$, but its
speed is smaller than that of the
$\sigma^{\overline{t}c\overline{\nu}\nu}$. For $m_{h_{t}}=250GeV
$, $ K^{tc}=0.05$, the ratio $R_{2}$ increases from 1.08 to 4.00
as $\sqrt{s}$ increases from 500GeV to 1500GeV. For
$\sqrt{s}=500GeV$ and in the range of $200GeV\leq m_{h_{t}}\leq
350 GeV$, $0.24 \leq R_{2}\leq 1.05$. Thus $h_{t}^{0}$ also can be
detected via the process
$e^{+}e^{-}\longrightarrow\gamma\overline{t}c $ at the LC
experiments.

   For $ m_{h_{t}}\leq 350 GeV$, $ gg $ is one of the dominant
decay modes of the top-Higgs $h_{t}^{0}$. $h_{t}^{0}$ may have
significantly contributions to the top-charm production at hadron
colliders. This has been extended studied by G. Burdman\cite{y4}.
His results show that the cross section of the gluon fusion
production and subsequent decay of the top-Higgs $h_{t}^{0}$ into
the $\overline{t}c $ final state is very large. There will be
several thousands of $ \overline{t}c $ events to be generated at
the LHC, which are larger than the number of
$\overline{t}c\overline{\nu}_{e}\nu_{e}$ events produced at the LC
experiments. Thus it is possible that the top-Higgs $h_{t}^{0}$
can be more easy detected at the LHC than at the LC experiments.
However, we must separate the signals from the large backgrounds
before observation of the top-Higgs $h_{t}^{0}$ at the LHC.

  For TC2 models, the underlying interactions, topcolor interactions,
are non-universal and therefore do not possess a GIM mechanism.
The non-universal interactions result in FCNC vertices when one
writes the interactions in the quark mass eigen-basis. Thus, the
top-Higgs $h_{t}^{0}$ can induce the new flavor changing scalar
coupling including the $t-c$ transitions. Considering the
production ratio of the FCNC process is negligible small in the
SM, we can use the FCNC process to discuss the observability of
the top-Higgs $h_{t}^{0}$. In this paper, we calculate the
contributions of $h_{t}^{0}$ to the processes
$e^{+}e^{-}\longrightarrow
\overline{t}c\overline{\nu}_{e}\nu_{e}$,
$e^{+}e^{-}\longrightarrow Z\overline{t}c$ and
$e^{+}e^{-}\longrightarrow \gamma\overline{t}c$ and compare the
results with each other at the  LC experiments with
$\sqrt{s}=500GeV-1500GeV$. Our results show that the processes
$e^{+}e^{-}\longrightarrow\overline{t}c\overline{\nu}_{e} \nu_{e}$
and $e^{+}e^{-}\longrightarrow \gamma\overline{t}c$ can be used to
detect the top-Higgs $h_{t}^{0}$ at LC the experiments with
$\sqrt{s}\geq 500GeV$. However, the process
$e^{+}e^{-}\longrightarrow\overline{t}c\overline{\nu}_{e}\nu_{e}$
is more sensitive to $h_{t}^{0}$ than that of
$e^{+}e^{-}\longrightarrow \gamma\overline {t}c$ , especially in
the LC experiments with $ \sqrt{s}\geq 1000GeV$.
\newpage
\begin{center}
{\bf Figure captions}
\end{center}
\begin{description}
\item[Fig.1:]The cross section
$\sigma^{\overline{t}c\overline{\nu}\nu}$ versus the top-Higgs
mass $m_{h_{t}}$ for the center-mass-energy $ \sqrt{s}=500$$GeV$
and $K^{tc}=0.02$, $0.05$, $0.08$, $0.1$.
\item[Fig.2:]The cross section
$\sigma^{\overline{t}c\overline{\nu}\nu}$ versus $m_{h_{t}}$ for $
K^{tc}=0.05 $ and $ \sqrt{s}=500GeV$, $1000GeV$, $1500GeV$.
\item[Fig.3:]The ratio $R_{1}=
\sigma^{\overline{t}c\overline{\nu}\nu}/ \sigma^{Z\overline{t}c}$
as a function of $m_{h_{t}}$ for $K^{tc}=0.05$ and $
\sqrt{s}=500GeV$, $1000GeV$, $1500GeV$.
\item[Fig.4:]The ratio $R_{2}=\sigma^{\overline{t}c\overline{\nu}\nu}/
\sigma^{\gamma\overline{t}c}$ as a function of $m_{h_{t}}$ for
$K^{tc}=0.05$ and $ \sqrt{s}=500GeV$, $1000GeV$, $1500GeV$.
\end{description}

 \newpage
\begin{figure}[pt]
\begin{center}
\begin{picture}(250,200)(0,0)
\put(-50,0){\epsfxsize120mm\epsfbox{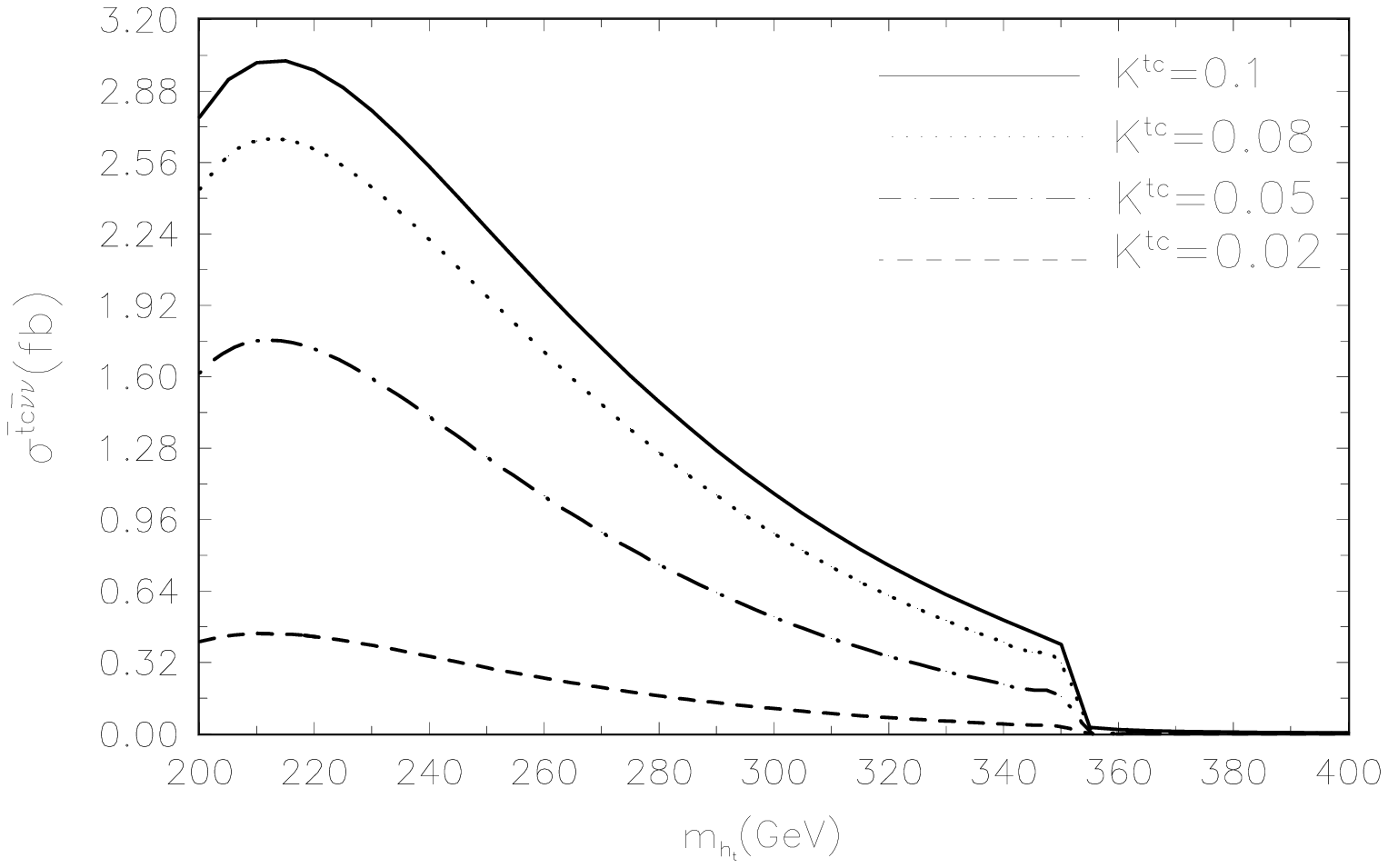}}
 \put(120,-10){ Fig.1}
\end{picture}
\end{center}
\end{figure}

\begin{figure}[hb]
\begin{center}
\begin{picture}(250,200)(0,0)
\put(-50,0){\epsfxsize120mm\epsfbox{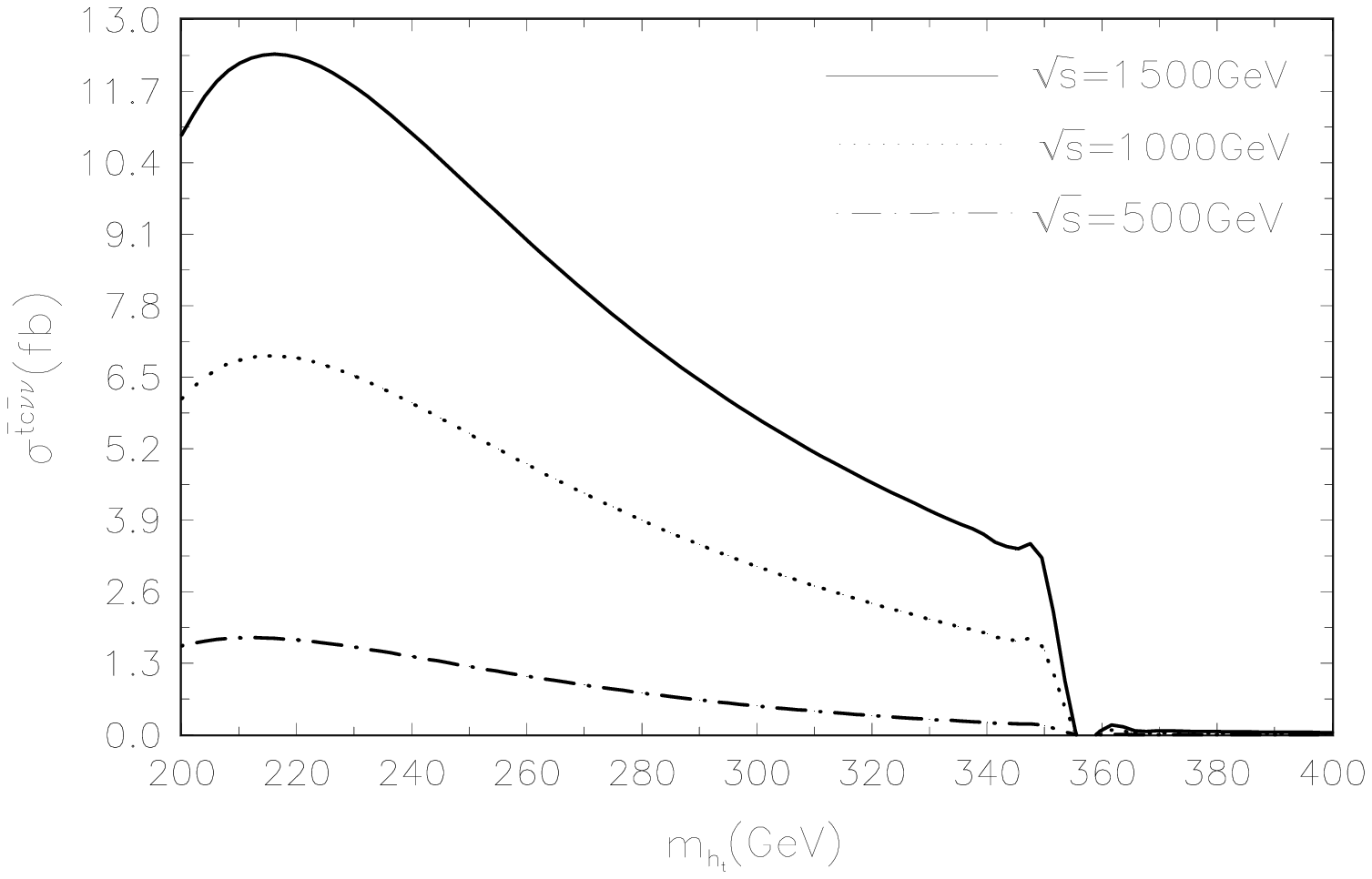}}
 \put(120,-10){ Fig.2}
\end{picture}
\end{center}
\end{figure}

\newpage
\begin{figure}[pt]
\begin{center}
\begin{picture}(250,200)(0,0)
\put(-50,0){\epsfxsize120mm\epsfbox{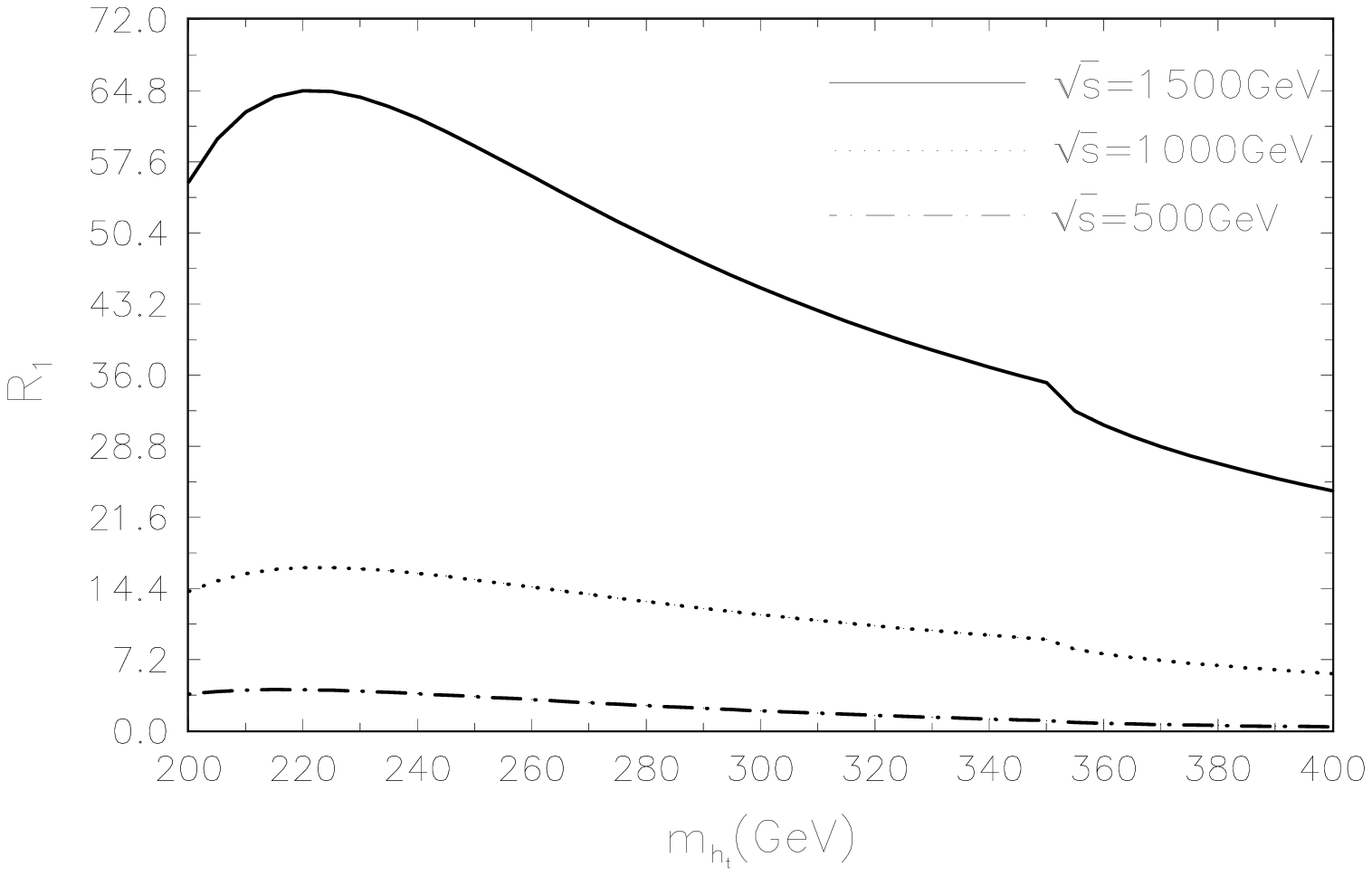}}
 \put(120,-10){ Fig.3}
\end{picture}
\end{center}
\end{figure}

\begin{figure}[hb]
\begin{center}
\begin{picture}(250,200)(0,0)
\put(-50,0){\epsfxsize120mm\epsfbox{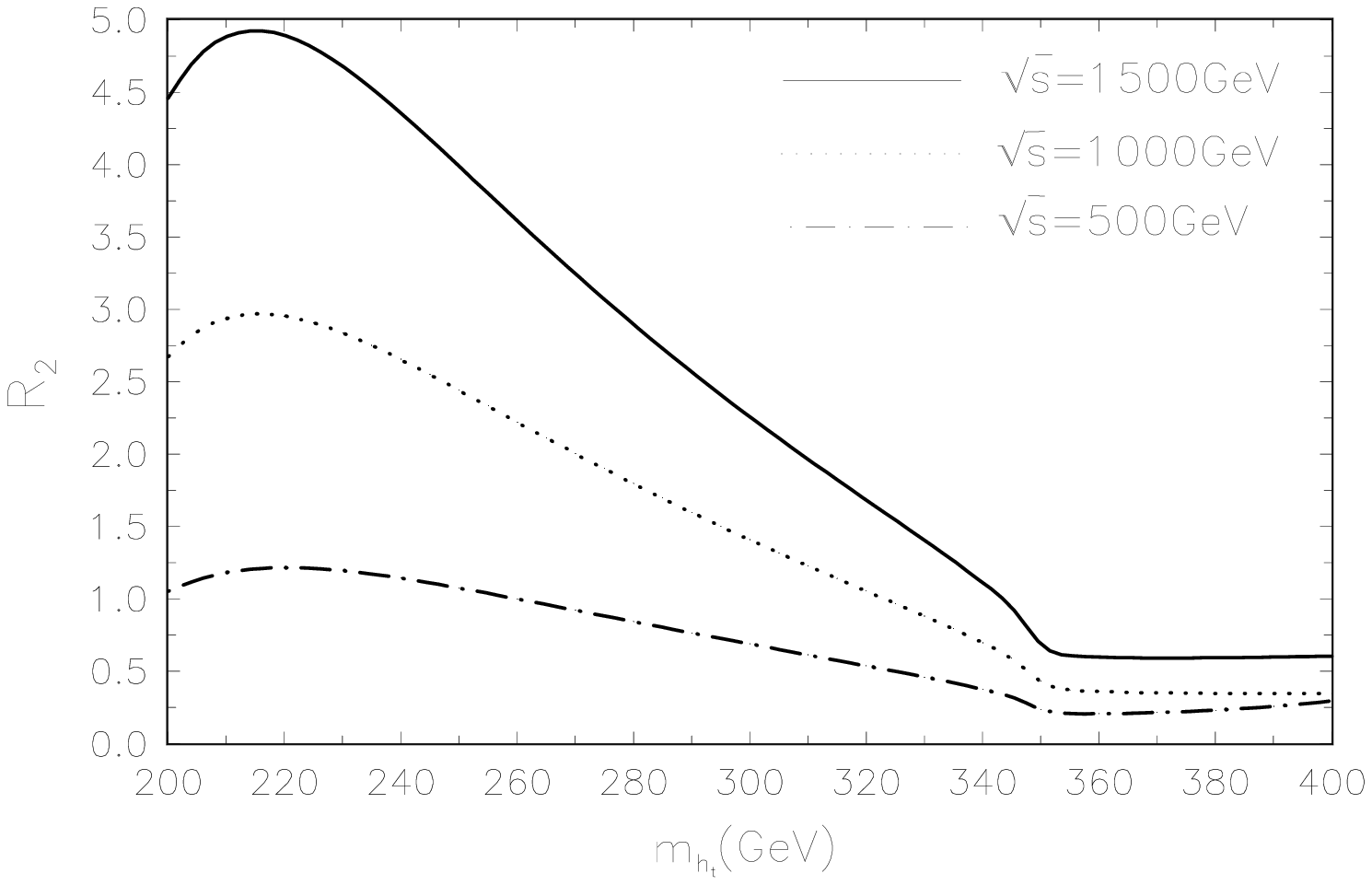}}
 \put(120,-10){ Fig.4}
\end{picture}
\end{center}
\end{figure}

\newpage
\begin{thebibliography}{99}
\bibitem{y1}C. T. Hill, {\em Phys.Lett. B}{\bf 345}(1995)483;
K. Lane and E. Eichten,
 {\em Phys.Lett. B}{\bf 352}(1995)383; K. Lane, {\em Phys.Lett.
 B}{\bf 433}(1998)96.
  \bibitem{y2}B.Dobrescu and C. T. Hill, {\em Phys.Rev.Lett.}{\bf 81}
  (1998)2634; R. S. Chivukula, B. Dobrescu, H. Georgi and C. T.
 Hill, {\em Phys.Rev. D}{\bf 59}(1999)075003.
  \bibitem{y3}M. B. Popovic and E. H. Simmons,
  {\em Phys.Rev. D}{\bf 58}(1998)095007;  K. Lane,  {\em Phys.Lett. B}{\bf 433}(1998)96; G.
 Burdman and N. Evans, {\em Phys.Rev. D}{\bf 59}(1999)115005.
  \bibitem{y4}Hong-Jian He and C.-P. Yuan, {\em
 Phys.Rev.Lett.} {\bf 83}(1999)28; G. Burdman, {\em
 Phys.Rev.Lett.} {\bf 83}(1999)2888.
\bibitem{y5}Chongxing Yue, et al., {\em Phys.Lett. B}{\bf 496}(2000)93;
Chongxing Yue, et al., hep-ph/0012332(to be published in Phys.
Rev.D).
\bibitem{y6}Chongxing Yue, et al., {\em Phys.Rev. D}{\bf
55}(1997)5541; Hangyi Zhou, Yuping Kuang, Chongxing Yue, Hua Wang,
{\em Nucl. Phys. Proc. Supp1. B}{\bf 75}(1999)302.
\bibitem{y7}M. Chanowitz and M.K. Gaillard,  {\em Phys.Lett.
B}{\bf 142}(1984)85; G. Kane, W. Repko and W. Rolick, {\em
Phys.Lett. B}{\bf 148}(1984)367; S. Dawson, {\em Nuel. Phys.
B}{\bf 249}(1985)42.
\bibitem{y8}S. Bar-Shalom, G. Eilam, A. Soni and J. Wudka,
{\em Phys.Rev.Lett.}{\bf 79}(1997)1217; {\em Phys.Rev. D}{\bf
57}(1998)2957.
 \end{thebibliography}
 \end{document}